# Best Practices for Administering Concept Inventories

Adrian Madsen, Sam McKagan
*American Association of Physics Teachers, College Park, MD*

Eleanor C. Sayre
*Department of Physics, Kansas State University, Manhattan, KS*

There is a plethora of concept inventories available for faculty to use, but it is not always clear exactly why you would use these tests, or how you should administer them and interpret the results. These research-based tests about physics and astronomy concepts are valuable because they allow for standardized comparisons among institutions, instructors, or over time. In order for these comparisons to be meaningful, you should use best practices for administering the tests. In interviews with 24 physics faculty[1], we have identified common questions that faculty members have about concept inventories. We have written this article to address common questions from interviews and provide a summary of best practices for administering concept inventories.

## Introduction to concept inventories

### What is a concept inventory?

Concept inventories are research-based assessment instruments that probe students' understanding of particular physics concepts[2,3]. We use the term "concept inventory" to refer to any kind of research-based assessment instrument that measures conceptual understanding, because this is how the term is most commonly used. Some researchers suggest more precise definitions for what a concept inventory is[2], but in this article we will use a more expansive definition which includes both multiple-choice and open-ended assessments that were developed following a rigorous research process (discussed below in the "What goes into the development of concept inventories?" section).

The most commonly used concept inventory is the Force Concept Inventory (FCI)[4], and there are over 60 other concept inventories for various introductory and upper-level topics in physics and astronomy[5]. Concept inventories are usually given at the beginning of a course (pre-test) to gain a sense of students' prior knowledge, and again at the end of the course (post-test) to gauge changes in their understanding.

Concept inventories are intended to measure the effectiveness of your teaching by assessing, on average, what your students learned in the course, and it follows that the results of the class as a whole are more important than individual students' scores[4]. Further, one study[6] suggests that concept inventories are not intended to be used as placement exams to place students into physics courses. Henderson[6] found that there is a non-zero number of students who have very low pre-test concept inventory scores but still earned an A in the course. A placement test is most effective if it can distinguish students who will succeed in the course and students who will not, and in this case, the concept inventory used did not do this.

Most concept inventories are multiple-choice, although there are several (approximately 4) upper-level concept inventories that are composed of all open-ended questions, and several (approximately 5) that are composed of a majority of multiple-choice questions and a couple of open-ended questions. The open-ended questions are usually graded with a rubric, and thus take more time to grade than multiple-choice questions, but can give you more insight into your students' understanding.

## Why should I use a concept inventory?

The detailed development of concept inventories makes them a unique form of assessment that speaks students' language and contains students' ideas. Faculty often think that the questions are too easy and are surprised when students do poorly[7]. This has been found to be especially true with the FCI[7], and the degree to which this happens with other concept inventories is less well known. The inclusion of students' everyday ideas and natural language in the multiple-choice options requires students to have a sophisticated understanding of the concept(s) in order to do well on these kinds of tests. For these reasons, concept inventories have had a major impact on physics education reform over the past 20 years. This started with the FCI, which was given to thousands of students throughout the country. The results show that research-based teaching methods, such as interactive engagement, lead to dramatic improvements in students' gains as a result of the course[8,9]. These results have inspired many physics instructors to give the FCI and other concept inventories to their students and to radically change their teaching methods from traditional lecture to interactive engagement techniques based on the results[7].

## What can I use a concept inventory for?

Concept inventories allow you to make comparisons of the effectiveness of your teaching over time and across instructors and institutions. They help you to answer questions such as, "How do my students' scores compare to other similar students?" or "I made a change to my teaching, did it work?" They have been rigorously designed and tested, so you know that (on average) your students' scores reflect their understanding of physics and not a misinterpretation of the wording of the questions. They are usually multiple-choice, so they are easy to score, but because of this format, students could guess the right answer without having a robust understanding of the concepts being tested. Because of this limitation, your students' scores on these tests give an upper bound for their understanding of the concepts tested. While it is possible for students to guess the right answer, instructors often find that their students score much lower than one would expect if they were guessing[10]. This is because the multiple-choice distractors are appealing to students since they are based on their own thinking and wording. Further, instructors use the *differences* in scores between pre- and post-test (e.g. raw gain, normalized gain or effect size), not the value of the scores themselves, to determine the effectiveness of their teaching. If scores were slightly inflated by guessing, looking at differences in scores cancels out this effect, and the comparisons remain valid. While high gains or high scores do not necessarily mean that students have learned "enough" or "a lot", low gains or low scores are pretty good indictors that they have not.

Concept inventories have many affordances, but are not meant to replace other assessments such as exams, homework, clicker questions, discussions with students, etc. Conventional wisdom is that they should be used in concert with these other forms of assessment to get a holistic picture of your students' learning and the effectiveness of your teaching. Concept inventories don't measure other important aspects of learning in your course, for example, your students' math skills, concepts covered in your course but not on the concept inventory, beliefs about learning physics, or problem-solving skills.

However, concept inventory questions can actually be much more difficult for students than typical end-of-chapter problems, because they require deep understanding. Many students solve end-of-chapter problems by rote application of algorithms without really understanding what they're doing. Students who do well on questions on concept inventories typically also do well on traditional problems, but the reverse is not necessarily true[10]. Further, there is evidence that focusing on the kind of basic conceptual understanding probed by concept inventories can lead to improvement on traditional problem-solving[7,10].

### What goes into the development of concept inventories?

Good concept inventories are different from typical exams in that their development involves extensive research and development by experts in physics education research to ensure that the questions represent concepts that faculty think are important, that the possible responses represent real student thinking and make sense to students, and that students' scores reliably tell us something about their understanding. The typical process of developing a concept inventory includes the following steps[3,10]:

1. Gathering students' ideas about a given topic, usually with interviews or open-ended written questions, and identifying patterns in these ideas.
2. Using students' ideas to develop questions where the responses cover the range of students' most common incorrect ideas using the students' actual wording.
3. Testing these questions with another group of students and ensuring that students choose the correct answer for the right reasons. Usually, researchers use interviews where students talk about their thinking for each question.
4. Testing these questions with experts in the discipline to ensure that they agree on the importance of the questions and the correctness of the answers.
5. Revising questions based on feedback from students and experts.
6. Administering concept inventory to large numbers of students. Checking the reproducibility of results across courses and institutions. Checking the distributions of answers. Using various statistical methods to ensure the reliability of the assessment.
7. Revising again.

This rigorous development process produces valid and reliable assessments that can be used to compare instruction across classes and institutions.

## Choosing a concept inventory

### Which concept inventory should I use?

There are over 60 different concept inventories in physics and astronomy, encompassing nearly every subject and many different levels. There are probably no concept inventories that exactly match the content of your course and the way that you teach the material, but using one that covers some of what you teach still allows you to get a big picture idea of the effectiveness of your teaching. The assessment pages on PhysPort.org[5] provide a full list of all 60+ concept inventories that you can filter by subject, level, format, and level of research validation. For most concept inventories, you can then download the test and an implementation guide that covers the specifics of using that concept inventory. You can also find information on the research and development process for each concept inventory. For areas where there are multiple concept inventories that cover similar content, such as introductory mechanics, see our resource letter[11] that compares them and gives recommendations for which to use.

### How do I evaluate the quality of a concept inventory?

It can be difficult for faculty members to evaluate the quality of a concept inventory because we don't think like a typical introductory physics student. The questions in a good concept inventory are based on research into the way students think, and are often designed to elicit common student ideas that are surprising to faculty. If you don't know about these ideas, you might think the questions are too easy or not understand the point. Further, not all concept inventories have the same level of rigorous research behind them. Some don't follow all the steps of research and development described above while some include extra steps that help make the assessment more appropriate for a particular purpose.

Based on the steps to developing a good research-based assessment (outlined above), we have created list of seven categories of research validation (Table I) in order to help faculty determine how rigorously a concept inventory has been developed.

TABLE I. Research validation categories

| |
|---|
| Questions based on research into student thinking |
| Studied with student interviews |
| Studied with expert review |
| Appropriate use of statistical analysis |
| Administered at multiple institutions |
| Research published by someone other than developers |
| At least one peer-reviewed publication |

We determine the level of research validation (Table II) for an assessment based on how many of the research validation categories apply to the concept inventory (Table I). The level of research validation as well as a summary of the research behind each assessment is given at PhysPort.org/assessments.

TABLE II. Determination of the level of research validation for an assessment.

| # Categories | Research validation level |
|---|---|
| All 7 | Gold |
| 5-6 | Silver |
| 3-4 | Bronze |
| 1-2 | Research-based |

## Administering concept inventories

### How do I administer a concept inventory?

**Pre-test and post-test:** For most concept inventories, you should give the test both as a pre-test and a post-test in order to measure student learning:

- Give the pre-test before covering any relevant course material so that you accurately capture students' incoming knowledge and not what they learned in the first few class sessions[12].
- Give the post-test at the end of the term.

For some concept inventories in areas such as quantum mechanics and electromagnetism, students are less likely to be familiar with the concepts before the class, so developers recommend using them only as a post-test. In these cases, pre-test scores are close to random guessing and students tend to find the tests demoralizing.

**In class or out of class:**
- Give the test in-class or in a supervised testing center. Many instrument developers strongly discourage giving concept inventories in an unsupervised manner outside of class because of students might use outside resources or share the test[10,13–17]. The classroom provides a standardized environment that usually results in a higher completion rate than if students take the test outside of class[18,19].

**Online or on paper:**
- Typically, concept inventories are given on paper in-class.
- To make grading more efficient, use scantrons or another similar tabulation system[20]. Often the testing services office at your university can give you an electronic spreadsheet of individual students' raw scores and overall summary information. Your students' raw responses allow you to upload your results to the PhysPort Data Explorer[21] and get automatic analysis of your results.
- It is more convenient to print the test question booklet and the answer page separately so that you can reuse test question booklets from section to section and year to year.
- Collect the answer sheets/scantrons and tests and store them securely.
- You can give the test online in a supervised testing environment such as during class (if computers are available), in a supervised testing center, or in your office. The supervised environment ensures test security, and giving it online makes it quicker and easier to analyze your results. For more information on giving concept inventories online, see our expert recommendation on PhysPort titled, "Guidelines for giving concept inventories online"[22].

**Ensuring validity:** To make comparisons with other classes meaningful:
- Use the whole test, with the original wording and question order.
- Give students the recommended time to take the test. You can find this on PhysPort by searching for your specific assessment[5].
- Look up any test-specific guidelines in the implementation guide for each specific assessment on PhysPort[5] and follow them. For example, some tests should only be given as post-tests.

**Encouraging participation:** For the results to most accurately represent students' real understanding, you should encourage your students to take the test seriously, but not provide so many incentives that they are tempted to look up the answers.
- Make the test required, and give credit for completing the test (but not for

correctness)[12]. (Incentives discussed further below).
- When you give the test to your students, test developers suggest giving it a generic title, like 'mechanics survey' to keep students from looking it up.
- Experts suggest that you stress that this assessment is designed to evaluate the <u>instructor</u> and the instructor's use of curriculum, not the strength of any individual student. Make it clear that their results will not influence their course grade. Let them know they are not expected to know the correct answers; you would like to know how they think about these questions. They should make their best guess if they are unsure about a question and should answer all the questions[3].

## What incentives should I give my students for taking the test?

A common concern when giving concept inventories is getting students to take the test seriously. Research[6] shows that a majority of students take concept inventories seriously when they are given in class and not graded. Henderson[6] compared students whose FCI score counted toward their final grade and those whose FCI score did not. He found evidence for no more than 2.8% of the students not taking the test seriously as a result of it not being graded. That said, there are many types of incentives you could use which have differing effects on completion rates and overall scores. Incentives discussed in the literature include:

- giving a small number of participation or extra credit points[3,10] (recommended)
- giving the post-test on the second-to-last day of class then reviewing for the final on the last day of class by discussing post-test questions that students did poorly on[3] (recommended[3], though some believe this compromises the security of the test[10])
- putting questions on the final exam[12] (**not** recommended)
- replacing lowest quiz grade if score 90% or higher on concept inventory[12] (**not** recommended)

Ding et al.[12] compared the completion rate and overall score on the post-test for a concept inventory using several of these incentives. Their findings are in Table III. Giving a small number of participation points increased participation rate while scores remained similar to giving no incentive. Adams et al.[3] and Redish[10] have had similar success giving a small number of participation points to help students take the test seriously. When the test became more high-stakes when it was given on the final exam, completion rates and scores increased. However, concept inventories are meant to assess the effectiveness of instruction and not individual students' concept mastery, so we do not recommend using concept inventories in this kind of high-stakes testing situation. Some instructors choose to give no incentives.

**Table III.** Comparison of completion rates and overall scores for various incentives given for the BEMA post-test[12].

|  | **Small number of participation points** | **Questions incorporated into final exam** | **Replace lowest quiz grade if score on concept inventory > 90%** |
|---|---|---|---|
| **No Incentive** | Higher completion rate for participation | Much higher completion rate and | Much lower participation rate and much higher |

| | points, similar scores. | higher scores for final exam. | scores for replacement of quiz grade. |

## What about test security?

Concept inventories are the result of a long and involved development process and are an extremely valuable tool for physics educators to improve their teaching. To help keep the tests secure, only verified educators are able to download more than 60+ physics and astronomy concept inventories on PhysPort.org[5]. Once you have access to a concept inventory, most test developers[10,13–17] ask that you ensure that students do not have unsupervised access to it either online or on paper. It's important that these tests are not available for students to review before they take them so that their scores are not artificially inflated and do not give you an inaccurate picture of the effectiveness of your teaching. After you have given the tests, it is recommended that you not post or hand out the tests or answers to your students. You are strongly discouraged from giving your students paper copies to complete at home.

# Scoring and interpreting your results

## How do I interpret my scores?

Once your students have completed the pre- and/or post-test, you need to figure out what your scores mean and how you can use them to improve your teaching.

*Pre-test Scores*

Pre-test scores give you a sense of your students' incoming knowledge. You can look at the average pre-test scores as well as scores on individual questions or question clusters[23] to determine the concepts your students understand well and where there are holes in their initial understanding. You can then adjust your instruction accordingly. For example, if your students seem to have a strong understanding of Newton's Second Law, you could spend less class time on this topic.

*Changes in Scores from Pre- to Post-test*

After your students have completed the pre- and post-tests, you can look at how their scores have changed as a result of your course. You can compute the average overall gain, normalized gain, and/or effect size using matched student data (only students who took both the pre- and post-test are included) and compare the results for your class to published results (Table IV). It is most common for physics faculty to look at the normalized gain for their results, while in social sciences research outside of physics, it is more common to report an effect size than a gain.

| Raw Gain | $<post> - <pre>$ |
|---|---|
| Normalized Gain | $\dfrac{<post> - <pre>}{100\% - <pre>}$ |

Table IV. Comparing raw gain, normalized gain & effect size

| Effect Size | $\dfrac{<post> - <pre>}{stdev}$ |
|---|---|

The raw gain compares the average post-test score to the average pre-test score. This is the crudest measure of how your students learn. The normalized gain compares the raw gain to how much your students didn't know at the beginning of the class. There are two ways typically used to calculate normalized gain:
- **Gain of averages:** First calculate the average pre-test and average post-test score for your class, then take the normalized gain of these: <g> = (<Post> - <Pre>)/(100 - <Pre>)
- **Average of gains:** First calculate the normalized gain for each student, then average these: $g_{ave}$ = <(Post - Pre)/(100 - <Pre>)>

According to Hake[8] and Bao[24] the difference between these two calculations is not significant for large classes, but may differ quite a bit for small classes. The gain of averages is the official definition given by Hake[8], but many researchers use the average of gains instead. For more details on the advantages and disadvantages of each, as well as other issues in calculating normalized gain, see our expert recommendation on normalized gain[25].

Gain does not account for the spread in students' scores. Effect size compares raw gain to the standard deviation of students' scores. Because the standard deviation includes how many students you have, using the effect size also lets you compare teaching effectiveness between classes of different sizes more fairly. For that reason, effect size is very popular among education researchers and statisticians. For more details, see our expert recommendation on effect size[26].

## How can I quickly and easily analyze my results? PhysPort Data Explorer

PhysPort offers a powerful and secure assessment Data Explorer that makes analyzing and interpreting your concept inventory results quick and easy. This online tool allows you to securely upload your students' concept inventory results and visualize them in a variety of ways. You can use the Data Explorer to calculate normalized gain and effect size quickly and compare your own assessment results over time as you make changes to your course. Many instructors who teach the same course several times find it extremely useful to document changes they made to their teaching and compare their gains over time to determine how the changes they made influenced their students' learning[27]. This kind of iterative process is exactly how concept inventories were intended to help improve instruction.

The Data Explorer also includes "one-click statistics" which performs appropriate statistical analyses on your assessment results and gives recommendations for your teaching based on your results. You can also look at how your students performed on individual questions or clusters[23] of post-test questions to get a coarse-grained sense of the concepts that students did not learn as well. You can then reflect on the way these concepts are taught in your course and look for ways to improve.

Additionally, you can download a report of your results and comparisons that you can use to talk to your colleagues about your course, include in tenure documents, accreditation reports, etc. The system has robust security measures to ensure that your students' assessment data and you and your students' identities are protected. Coming soon, the Data Explorer will enable you to compare your assessment results to other "students like yours" from similar institutions and in similar courses. To use the data explorer go to www.physport.org/dataexplorer.

### How does teaching to the test impact my scores?

RBAs are one way to measure teaching improvement, but a common concern is that instructors are "teaching to the test" so their results don't represent the effectiveness of their teaching. Test developers recommend that you don't go over the exact test questions in class before the test[10], as you may be testing students ability to recall information they have previously seen instead of applying their conceptual understanding to a new situation.

Another way to think about "teaching to the test" is teaching content that aligns well with what is being tested (but not going over the exact test questions). It could be that larger RBA gains are the result of teaching improvement or teaching that is better tuned to the test. In the latter case, these gains wouldn't mean that your students are necessarily learning more; they may just be learning content that is better measured by your chosen test. Hake[8] explains, "In the broadest sense, IE (interactive engagement) courses all ''teach to the test'' to some extent if this means teaching so as to give students some understanding of the basic concepts of Newtonian mechanics as examined on the FCI/MD tests. However, this is the bias we are attempting to measure." Research suggests additional lecture time on the topic being tested does not result in better RBA scores. In a study by Redish, Saul, and Steinberg[28], the same award-winning professor taught two sections of the same class. In one section he gave three hours of lectures on the same topic "teaching to the test" without actually going over the test questions, followed by a traditional recitation on the topic. In the other section he gave only one hour of lecture and they did a tutorial in recitation section, and the students performed better on the FCI.

In contrast, you could be teaching content very effectively that doesn't align with what is being tested. For example, the Matters and Interactions (M&I) curriculum focuses on "the generality of fundamental physical principles, the introduction of microscopic models of matter, and its coherence in linking different domains of physics[29]." FCI gains were compared for courses taught with a traditional mechanics curriculum and M&I curriculum, both using similar interactive pedagogy. Courses taught with the traditional mechanics curriculum had higher gains that could be explained by the difference in the fraction of homework and course time spent on force and motion topics tested on the FCI.

When you are trying to interpret RBA results, consider three cases. First, avoid going over exact test questions with your students before giving then the test. Second, you could be teaching important content and doing it well, and the RBA you are using is not reflecting this. Third, you might be teaching content that does align well with your RBA, and teaching it very effectively (or not), and the test does reflect this.

### Do some kinds of students tend to do better than other kinds of students?

Studies have looked at various characteristics of students, including gender[30], previous preparation[9], and ethnicity[31,32], to investigate how these relate to their scores on commonly used RBAs. It is important to look for differences in scores based on characteristics of students, as there are systematic biases in our educational system and society at large that could differentially influence certain students' ability to score well on these tests. If this is the case, instructors should be aware of these differences and be sure not to put too much emphasis on assessments that are known to advantage one particular group of students.

In an analysis of 9 studies of FCI and FMCE scores[30], we found that men's average pretest scores are always higher than women's, and in most cases men's posttest scores and normalized gains are higher as well. There is sometimes a similar gender gap on the BEMA and CSEM, but it is usually much smaller and sometimes is zero or favors women. We also investigated 30 factors that might explain this gender gap, and found that no single factor is sufficient to explain the gap. There are some factors that do contribute to a difference between male and female responses, but the size of these differences is smaller than the size of the overall gender gap, suggesting that the gender gap is most likely due to the combination of many small factors (systematic biases) rather than any one factor that can easily be modified.

There have been two studies[31,32] looking at how ethnicity influences FCI scores. They both find a gap in pre- and post-test scores based on ethnicity, but the gap disappeared when you looked at raw gain for one study[31], where as a gap in raw gain and normalized gain was found in the other study[32]. So there are differences in FCI scores based on ethnicity, but the nature of this difference is not well understood, so it should interpreted with caution.

Von Korff et al.[9] performed a meta-analysis of FCI and FMCE data from 72 papers, representing about 600 classes and about 45000 students. They found that neither institutional average SAT scores nor class average pre-test score correlated with class average normalized gains. This implies that prior preparation of students is not influencing normalized gains on these tests. They also did *not* find a correlation between institution type (whether it is an Associates-, Bachelors-, Masters-, or Doctoral-granting) and normalized gains, but they did find a correlation between pre-test scores and institution type. This suggests there are differences in *prior preparation* based on type of institution, but these differences do not influence normalized gains. Overall, these results are reassuring, as they imply that we can compare normalized gain scores for students from different types of institutions (but not pre-test scores). But, this study was limited by selection effects in publishing and only being able to look at class averages rather than individual students.

Overall, RBA scores can tell you about the effectiveness of your teaching when you look at the aggregate results for your entire class. Further, when looking at class level results, you can compare normalized gains to other institutions that aren't like yours. Instructors should be cautious about looking at scores for individual students as there is evidence pointing to systematic biases that disadvantage certain students.

# Conclusion

We have written this article to address common questions about giving concept inventories and provided a summary of best practices for administering concept inventories. For the answers to even more questions about concept inventories see our expert recommendation "Addressing Common Concerns about Concept Inventories"[33] on PhysPort.org. You can also find more information on normalized gain[25], effect size[26], giving concept inventories online[22], and getting your students' answers from concept inventories into spreadsheets[34].

**Endnotes**